\documentclass[pra,10pt,twocolumn,amsmath,floatfix,showpacs]{revtex4}
\usepackage{times}
\usepackage{mathptmx}
\usepackage{graphicx}
\usepackage{amsmath}

\newcommand{\parti}[2]{\frac{\partial #1}{\partial #2}}
\newcommand{\partit}[2]{\frac{\partial^2 #1}{\partial #2^2}}
\newcommand{\diff}[2]{\frac{d #1}{d #2}}

\newcommand{\ket}[1]{|#1\rangle}
\newcommand{\bra}[1]{\langle#1|}
\newcommand{\avg}[1]{\langle#1\rangle}
\newcommand{\Avg}[1]{\left\langle#1\right\rangle}
\newcommand{\sech}{\operatorname{sech}}
\begin{document}
\title{Decoherence of Quantum-Enhanced Timing Accuracy}

\author{Mankei Tsang}
\date{\today}
\affiliation{Department of Electrical Engineering, 
California Institute of Technology, Pasadena, CA 91125}

\begin{abstract}
  Quantum enhancement of optical pulse timing accuracy is investigated
  in the Heisenberg picture. Effects of optical loss, group-velocity
  dispersion, and Kerr nonlinearity on the position and momentum of an
  optical pulse are studied via Heisenberg equations of motion.  Using
  the developed formalism, the impact of decoherence by optical loss
  on the use of adiabatic soliton control for beating the timing
  standard quantum limit [Tsang, \prl \textbf{97}, 023902 (2006)] is
  analyzed theoretically and numerically.  The analysis shows that an
  appreciable enhancement can be achieved using current technology,
  despite an increase in timing jitter mainly due to the Gordon-Haus
  effect.  The decoherence effect of optical loss on the transmission
  of quantum-enhanced timing information is also studied, in order to
  identify situations in which the enhancement is able to survive.
\end{abstract}

\maketitle

\section{Introduction}
It has been suggested that the use of correlated photons is able to
enhance the position accuracy of an optical pulse beyond the standard
quantum limit, and the enhancement can be useful for positioning and
clock synchronization applications \cite{giovannetti_nature}.
Generation of two photons with the requisite correlation has been
demonstrated experimentally by Kuzucu \textit{et al.}\ \cite{kuzucu},
but in practice it is more desirable to produce as many correlated
photons as possible in order to obtain a higher accuracy. To achieve
quantum enhancement for a large number of photons, a scheme of
adiabatically manipulating optical fiber solitons has recently been
proposed \cite{tsang_prl}, opening up a viable route of applying
quantum enhancement to practical situations. The analysis in
Ref.~\cite{tsang_prl} assumes that the optical fibers are lossless, so
the Heisenberg limit \cite{giovannetti_sci} can be reached in
principle. In reality, however, the quantum noise associated with
optical loss increases the soliton timing jitter and limits the
achievable enhancement.  Compared with the use of solitons for
quadrature squeezing \cite{carter}, the adiabatic
soliton control scheme potentially suffers more from decoherence,
because the soliton must propagate for a longer distance to satisfy
the adiabatic approximation.  The effect of loss on a similar scheme
of soliton momentum squeezing has been studied by Fini and Hagelstein
\cite{fini}, although they did not study the timing jitter evolution
relevant to the scheme in Ref.~\cite{tsang_prl}, and did not take into
account possible departure from the adiabatic approximation.

In this paper, the decoherence effect of optical loss on the timing
accuracy enhancement scheme proposed in Ref.~\cite{tsang_prl} is
investigated in depth, in order to evaluate the performance of the
scheme in practice.  Instead of approaching the problem in the
Schr\"odinger picture like prior work
\cite{giovannetti_nature,fini,fini_config,tsang_prl}, this paper
primarily utilizes Heisenberg equations of motion, since they are able
to account for dissipation and fluctuation in a more elegant way. For
simplicity, scalar solitons, as opposed to vector solitons studied in
Ref.~\cite{tsang_prl}, are considered here. The theoretical and
numerical analyses show that, despite an increase in timing jitter due
to quantum noise and deviation from the adiabatic approximation, an
appreciable enhancement can still be achieved using a realistic setup.

The developed formalism is also used to study the propagation of an
optical pulse with quantum-enhanced timing accuracy in a lossy,
dispersive, and nonlinear medium, such as an optical fiber, in order
to identify situations in which the enhancement can still survive.
The effect of loss on many correlated photons sent in as many channels has
been investigated by Giovannetti \textit{et al.}\
\cite{giovannetti_nature}, but their analysis focuses on a relatively
small number of correlated photons and does not include the effects of
dispersion and nonlinearity.

This paper is organized as follows: Section \ref{def} defines the
general theoretical framework, and derives the standard quantum limits
and Heisenberg limits on the variances of the pulse position and
momentum operators.  Section \ref{heisenberg} studies the evolution of
such operators in the presence of loss, group-velocity dispersion, and
Kerr nonlinearity, and determines the effect of dissipation and
fluctuation on the position and momentum uncertainties. Section
\ref{soliton} investigates theoretically and numerically the impact of
optical loss on the adiabatic soliton control scheme using realistic
parameters, while Sec.~\ref{transmission} studies the decoherence
effect on the transmission of the quantum-enhanced timing information
in various linear and nonlinear systems.

\section{\label{def}Theoretical Framework}
\subsection{Definition of pulse position and momentum operators}
The positive-frequency electric field of a waveguide mode at a certain
longitudinal position can be defined as \cite{huttner}
\begin{align}
\hat{E}^{(+)}(t) &= i\int_0^\infty d\omega
\left(\frac{\hbar\omega\eta}
{4\pi\epsilon_0cn^2S}\right)^{1/2}
\hat{c}(\omega)e^{-i\omega t},
\end{align}
where $n$ is the refractive index, $\eta$ is the real part of $n$, $S$
is the transverse area of the waveguide mode, and $\hat{c}(\omega)$ is
the photon annihilation operator. The annihilation operator is related
to the corresponding creation operator via the commutator
\cite{huttner},
\begin{align}
[\hat{c}(\omega),\hat{c}^\dagger(\omega')] &= \delta(\omega-\omega').
\end{align}
For a pulse with a slowly-varying envelope compared with the optical
frequency, the coefficient in front of the annihilation operator can
be assumed to be independent of frequency and can be evaluated at the
carrier frequency $\omega_0$, so that the electric field is
proportional to the temporal envelope annihilation operator
$\hat{A}(t)$,
\begin{align}
\hat{E}^{(+)}(t) &\propto
\hat{A}(t)e^{-i\omega_0 t},
\\
\hat{A}(t) &\equiv
\frac{1}{\sqrt{2\pi}}\int d\omega\ \hat{a}(\omega)e^{-i\omega t},
\\
\hat{a}(\omega) &\equiv \hat{c}(\omega+\omega_0).
\end{align}
The temporal envelope operator $\hat{A}(t)$ and the spectral operator
$\hat{a}(\omega)$ evidently also satisfy the following commutation
relations with their corresponding creation operators,
\begin{align}
[\hat{A}(t),\hat{A}^\dagger(t')] &= \delta(t-t'),
\\
[\hat{a}(\omega),\hat{a}^\dagger(\omega')] &= \delta(\omega-\omega').
\end{align}
The total photon number operator can be defined as
\begin{align}
\hat{N} &\equiv \int dt\ \hat{A}^\dagger(t)\hat{A}(t),
\label{number}
\end{align}
and the pulse center position operator as \cite{vaughan}
\begin{align}
\hat{T} &\equiv \frac{1}{N} \int dt\ 
t\hat{A}^\dagger(t)\hat{A}(t),
\label{position}
\end{align}
where 
\begin{align}
N \equiv \Avg{\hat{N}}
\end{align}
is the average photon number. This definition uses $1/N$ as the
normalization coefficient, instead of the inverse photon number
operator $\hat{N}^{-1}$ used by Lai and Haus \cite{lai1}, in order to
express the position operator in terms of normally ordered optical
field operators that are easier to handle, as well as to avoid the
potential problem of applying $\hat{N}^{-1}$ on the vacuum state.  As
long as the photon-number fluctuation is small, the position operator
naturally corresponds to the measurement of the center position of the
pulse intensity profile.  An average longitudinal momentum operator
can be similarly defined,
\begin{align}
\hat{\Omega} &\equiv \frac{1}{N}
\int d\omega\ \omega\hat{a}^\dagger(\omega)
\hat{a}(\omega)
\nonumber\\
&=\frac{1}{N} \int dt\
\hat{A}^\dagger(t)\left(i\parti{}{t}\right)\hat{A}(t).
\label{momentum}
\end{align}
If the quantum state is close to a large-photon-number coherent state,
$\hat{A}$ can be approximated as $\Avg{\hat{A}} +\delta \hat{A}$, with
$O(\delta\hat{A})\ll O(\hat{A})$.  Equations (\ref{position}) and
(\ref{momentum}) then become the approximate position and momentum
operators defined by Haus and Lai for solitons in a linearized
approach \cite{haus_josab}.  The linearized expressions also describe
how they can be accurately measured in practice using balanced
homodyne detection \cite{haus_josab}.

For simplicity, we shall hereafter assume that $\Avg{\hat{T}} = 0$ and
$\Avg{\hat{\Omega}} = 0$ \cite{vaughan}.  In the systems considered in
this paper, these two quantities remain constant throughout
propagation, if $t$ is regarded as the retarded time in the moving
frame of the optical pulse.

The commutator between the position and momentum operators is
\begin{align}
[\hat{T},\hat{\Omega}] &= \frac{i\hat{N}}{N^2}.
\end{align}
By the Heisenberg uncertainty principle, 
\begin{align}
\Avg{\hat{T}^2}\Avg{\hat\Omega^2} \ge 
\left(\frac{1}{2i}\Avg{[\hat{T},\hat{\Omega}]}\right)^2
= \frac{1}{4N^2}.
\label{principle}
\end{align}

\subsection{Derivation of standard quantum limits}
The standard quantum limits and Heisenberg limits on
$\Avg{\hat{\Omega}^2}$ and $\Avg{\hat{T}^2}$ should be expressed in
terms of the pulse width $\Delta t$, defined as
\begin{align}
\Delta t &\equiv \Avg{\frac{1}{N}\int dt\
t^2 \hat{A}^\dagger(t)\hat{A}(t)}^{1/2},
\label{pulsewidth}
\end{align}
and the bandwidth $\Delta\omega$,
\begin{align}
\Delta\omega &\equiv \Avg{\frac{1}{N}\int d\omega\ \omega^2
\hat{a}^\dagger(\omega)\hat{a}(\omega)}^{1/2}
\nonumber\\
&=\Avg{\frac{1}{N}\int dt\ \hat{A}^\dagger(t)
\left(-\partit{}{t}\right)\hat{A}(t)}^{1/2}.
\label{bandwidth}
\end{align}
To calculate the standard quantum limit on the position uncertainty,
consider the expansion
\begin{align}
\Avg{\hat{\Omega}^2} &= \frac{1}{N^2}
\Avg{\int d\omega\ \omega \hat{a}^\dagger \hat{a} 
\int d\omega'\ \omega'\hat{a}'^\dagger\hat{a}'},
\end{align}
where we have written $\hat{a}=\hat{a}(\omega)$ and
$\hat{a}'=\hat{a}(\omega')$ as shorthands. Rearranging the
operators,
\begin{align}
\Avg{\hat{\Omega}^2}
&=\frac{1}{N}\Avg{\frac{1}{N}
\int d\omega\ \omega^2\hat{a}^\dagger \hat{a} }
\nonumber\\&\quad
+\frac{1}{N^2}\Avg{ \int d\omega \int d\omega'\ 
\omega\omega'\hat{a}^\dagger \hat{a}'^\dagger\hat{a}\hat{a}'}.
\label{expansion}
\end{align}
The first term on the right-hand side of Eq.~(\ref{expansion}) is
proportional to $\Delta\omega^2$, while the second term contains a
normally ordered cross-spectral density.  To derive the standard
quantum limit, we shall assume that the cross-spectral density
satisfies the factorization condition:
\begin{align}
\Avg{\hat{a}^\dagger\hat{a}'^\dagger
\hat{a}\hat{a}'} &\propto
\Avg{\hat{a}^\dagger\hat{a}}
\Avg{\hat{a}'^\dagger\hat{a}'}.
\label{factorize}
\end{align}
This condition is always satisfied by any pure or mixed state with
only one excited optical mode, such as a coherent state
\cite{titulaer,mandel}.  The second term on the right-hand side of
Eq.~(\ref{expansion}) becomes
\begin{align}
\frac{1}{N^2}\int d\omega \int d\omega'\ 
\omega\omega'\Avg{\hat{a}^\dagger \hat{a}'^\dagger\hat{a}\hat{a}'}
\propto \Avg{\hat{\Omega}}^2,
\end{align}
which is assumed to be zero, as per the convention of this paper.
Thus, the variance of $\hat{\Omega}$ is
\begin{align}
\Avg{\hat{\Omega}^2}_{\textrm{coh}} &= \frac{\Delta\omega^2}{N},
\label{VarOmega_CF}
\end{align}
where the subscript ``coh'' denotes statistics of coherent fields
\cite{titulaer,mandel} given by Eq.~(\ref{factorize}).  By virtue of
the Heisenberg uncertainty principle given by Eq.~(\ref{principle}),
the standard quantum limit on the position variance is hence
\begin{align}
\Avg{\hat{T}^2} &\ge \Avg{\hat{T}^2}_{\textrm{SQL}}=
\frac{1}{4N^2\Avg{\hat{\Omega}^2}_{\textrm{coh}}}=
\frac{1}{4N\Delta\omega^2}.
\label{position_sql}
\end{align}
This limit is applicable to any pure or mixed state, and is consistent
with the one suggested by Giovannetti \textit{et al.}\ for Fock states
\cite{giovannetti_nature}. A very similar derivation of the limit for
Fock states and coherent states is also performed by Vaughan
\textit{et al.}\ \cite{vaughan}.

Owing to Fourier duality of position and momentum in the
slowly-varying envelope regime, the standard quantum limit on
the momentum can be derived in the same way. The variance of
$\hat{T}$, assuming coherent-field statistics, is
\begin{align}
\Avg{\hat{T}^2}_{\textrm{coh}} &= \frac{\Delta t^2}{N},
\label{VarT_CF}
\end{align}
and the standard quantum limit on the momentum variance is
\begin{align}
\Avg{\hat{\Omega}^2}_{\textrm{SQL}} &= \frac{1}{4N^2\Avg{\hat{T}^2}_{\textrm{coh}}}
=\frac{1}{4N\Delta t^2}.
\label{momentum_sql}
\end{align}

\subsection{Derivation of Heisenberg limits}
To derive the Heisenberg limit on the position uncertainty, one needs
an absolute upper bound on the momentum uncertainty
$\Avg{\hat{\Omega}^2}$.  Consider the following non-negative quantity
proportional to the coherence bandwidth squared,
\begin{align}
\int d\omega \int d\omega'\
(\omega-\omega')^2
\Avg{\hat{a}^\dagger\hat{a}'^\dagger\hat{a}\hat{a}'} \ge 0.
\end{align}
This quantity is non-negative because $(\omega-\omega')^2$ is
non-negative and
$\Avg{\hat{a}^\dagger\hat{a}'^\dagger\hat{a}\hat{a}'}$ is also
non-negative \cite{mandel}.  It can be rewritten as
\begin{align}
&\quad \int d\omega \int d\omega'\
(\omega-\omega')^2\Avg{\hat{a}^\dagger\hat{a}'^\dagger\hat{a}\hat{a}'}
\nonumber\\
&=\int d\omega \int d\omega'\
(\omega-\omega')^2\Avg{\hat{a}^\dagger\hat{a}\hat{a}'^\dagger\hat{a}'},
\end{align}
and expanded as
\begin{align}
\Avg{\int d\omega \int d\omega'\ (\omega^2+\omega'^2-2\omega\omega')
\hat{a}^\dagger\hat{a}\hat{a}'^\dagger\hat{a}'}\ge 0,
\nonumber\\
2\Avg{\hat{N}\int d\omega\ \omega^2\hat{a}^\dagger\hat{a}}-
2N^2\Avg{\hat{\Omega}^2} \ge 0.
\end{align}
Here we shall approximate $\hat{N}$ with $N$, and neglect any
photon-number fluctuation.  This approximation is exact for Fock
states, and acceptable for any quantum state with a small
photon-number fluctuation, such as a large-photon-number coherent
state. We then obtain the following approximate inequality,
\begin{align}
\Avg{\hat{\Omega}^2} \le \Delta\omega^2.
\label{approx_ineq}
\end{align}
With the Heisenberg uncertainty principle given by
Eq.~(\ref{principle}) and the upper bound on $\Avg{\hat\Omega^2}$
given by Eq.~(\ref{approx_ineq}), one can then obtain the Heisenberg
limit on the uncertainty of $\hat{T}$:
\begin{align}
\Avg{\hat{T}^2} \ge \Avg{\hat{T}^2}_{\textrm{H}} = \frac{1}{4N^2\Delta\omega^2}.
\label{h_limit}
\end{align}
Equation (\ref{h_limit}) is again consistent with the Heisenberg limit
suggested by Giovannetti \textit{et al.}\ \cite{giovannetti_nature},
although the derivation here shows that it is not only valid for Fock
states but also correct to the first order for any quantum state with
a small photon-number fluctuation.

The Heisenberg limit on $\Avg{\hat\Omega^2}$ is similar,
\begin{align}
\Avg{\hat{\Omega}^2}_{\textrm{H}} &= \frac{1}{4N^2\Delta t^2}.
\end{align}
A more exact derivation of the Heisenberg limits is given in Appendix
\ref{exact_heisenberg}, where the inverse photon-number operator
$\hat{N}^{-1}$ is used instead of $1/N$ in the definitions of
$\hat{T}$, $\hat{\Omega}$, $\Delta\omega$, and $\Delta t$. The
difference between the approximate Heisenberg limits derived here and
the exact Heisenberg limits in Appendix \ref{exact_heisenberg} is
negligible for small photon-number fluctuations.

\section{\label{heisenberg}Optical Pulse Propagation in the Heisenberg
  Picture}
The classical nonlinear Schr\"odinger equation that describes the
propagation of pulses in a lossy, dispersive, and nonlinear medium,
such as an optical fiber, is given by \cite{agrawal}
\begin{align}
i\parti{A}{z} &=\frac{\beta}{2}\partit{A}{t} - \kappa|A|^2A
-\frac{i\alpha}{2}A,
\label{cnse}
\end{align}
where $t$ is the retarded time coordinate in the frame of the moving
pulse, $\beta$ is the group-velocity dispersion coefficient, $\kappa$
is the normalized Kerr coefficient, and $\alpha$ is the loss
coefficient, all of which may depend on $z$.  The phenomenological
quantized version that preserves the commutator between $\hat{A}$ and
$\hat{A}^\dagger$ is \cite{carter}
\begin{align}
i\parti{\hat{A}}{z} &= \frac{\beta}{2}\partit{\hat{A}}{t}
- \kappa\hat{A}^\dagger\hat{A}\hat{A}
-\frac{i\alpha}{2}\hat{A} + i\hat{s}.
\label{qnse}
\end{align}
$\hat{A} \equiv \hat{A}(z,t)$ is the pulse envelope annihilation
operator in the Heisenberg picture, and $\hat{s}$ is the Langevin
noise operator, satisfying the commutation relation
\begin{align}
[\hat{s}(z,t),\hat{s}^\dagger(z',t')] &= \alpha\delta(z-z')\delta(t-t').
\end{align}
Rewriting the position and momentum operators in
Eqs.~(\ref{position}) and (\ref{momentum}) in the Heisenberg picture
as $\hat{T}(z)$ and $\hat{\Omega}(z)$ in terms of $\hat{A}(z,t)$,
differenting them with respect to $z$, and
using Eq.~(\ref{qnse}), their equations of motion can be derived,
\begin{align}
\diff{\hat{T}(z)}{z} &=\beta(z)\hat{\Omega}(z)+\hat{S}_T(z),
\\
\diff{\hat{\Omega}(z)}{z} &= \hat{S}_\Omega(z),
\end{align}
where $\hat{S}_T$ and $\hat{S}_\Omega$ are
position and momentum noise operators defined as
\begin{align}
\hat{S}_T(z) &\equiv
\frac{1}{N(z)}\int dt\ t\hat{s}^\dagger(z,t) \hat{A}(z,t) +
\textrm{H.\ c.},
\label{S_T}
\\
\hat{S}_\Omega(z) &\equiv
\frac{1}{N(z)}\int dt\ \hat{s}^\dagger(z,t)\left(i\parti{}{t}\right)
\hat{A}(z,t)+\textrm{H.\ c.},
\label{S_Omega}
\end{align}
and H.\ c.\ denotes Hermitian conjugate.  If the noise reservoir is
assumed to be in the vacuum state, the noise operators have the
following statistical properties, as shown in Appendix \ref{langevin},
\begin{align}
\Avg{\hat{S}_T(z)} &= 0,\
\Avg{\hat{S}_\Omega(z)} = 0,
\\
\Avg{\hat{S}_T(z)\hat{S}_T(z')} &=
\frac{\alpha(z)\Delta t^2(z)}{N(z)}\delta(z-z'),
\label{position_noise}\\
\Avg{\hat{S}_\Omega(z)\hat{S}_\Omega(z')} &=
\frac{\alpha(z)\Delta \omega^2(z)}{N(z)}\delta(z-z'),
\label{momentum_noise}\\
\Avg{\hat{S}_T(z)\hat{S}_\Omega(z')+\hat{S}_\Omega(z)\hat{S}_T(z')}
&=\frac{\alpha(z)C(z)}{N(z)}\delta(z-z'),
\label{cross_correlation}
\end{align}
where $C(z)$ is the pulse chirp factor, defined as
\begin{align}
C(z) &\equiv \Avg{\frac{1}{N(z)}\int dt\ 
\hat{A}^\dagger(z,t)\left[t\left(i\parti{}{t}\right)
+\left(i\parti{}{t}\right)t\right]\hat{A}(z,t)}.
\end{align}
The average position $\avg{\hat{T}(z)}$ and average momentum
$\avg{\hat{\Omega}(z)}$ are constant and assumed to be zero throughout
propagation. The variance of $\hat{\Omega}$ is then
\begin{align}
\Avg{\hat{\Omega}^2(z)} &= \Avg{\hat{\Omega}^2(0)}+
\int_0^z dz'\frac{\alpha(z')\Delta\omega^2(z')}{N(z')},
\label{freq_jitter}
\end{align}
while the variance of $\hat{T}$ is more complicated
due to the presence of dispersion,
\begin{align}
\Avg{\hat{T}^2(z)} &= \Avg{\hat{T}^2(0)}
+\Avg{\hat{T}(0)\hat{\Omega}(0)+\hat{\Omega}(0)\hat{T}(0)}
\int_0^z dz'\beta(z')
\nonumber\\&\quad
+\Avg{\hat{\Omega}^2(0)}\left[\int_0^z dz'\beta(z')\right]^2
\nonumber\\&\quad
+\int_0^z dz'\frac{\alpha(z')\Delta t^2(z')}{N(z')}
\nonumber\\&\quad
+\int_0^z dz'\beta(z')\int_0^{z'} dz''\frac{\alpha(z'')C(z'')}{N(z'')}
\nonumber\\&\quad
+2\int_0^z dz'\beta(z')\int_0^{z'} dz''\beta(z'')
\nonumber\\&\quad\times
\int_0^{z''}dz'''\frac{\alpha(z''')\Delta\omega^2(z''')}{N(z''')}.
\label{jitter}
\end{align}
Equation (\ref{jitter}) is the central result of this paper.  It is
similar to that derived by Haus for optical solitons using a
linearized approach \cite{haus_josab2}, but Eq.~(\ref{jitter}) is
valid for arbitrary loss, arbitrary dispersion profile $\beta(z)$, and
arbitrary evolution of pulse width $\Delta t(z)$, chirp $C(z)$, and
bandwidth $\Delta\omega(z)$, so that it is able to describe the effect
of loss on the quantum enhancement scheme proposed in
Ref.~\cite{tsang_prl}.  The first term on the right-hand side of
Eq.~(\ref{jitter}) is the initial quantum fluctuation, while the
second and third term on the right-hand side describe the quantum
dispersion effect \cite{lai2}.  In an ideal scenario described in
Ref.~\cite{tsang_prl}, $\Avg{\hat{T}^2(z)}$ remains constant if the
net dispersion $\int_0^zdz'\beta(z')$ is zero and quantum dispersion
is compensated.  With loss, however, noise introduces a diffusive
jitter given by the fourth term on the right-hand side of
Eq.~(\ref{jitter}),
\begin{align}
\Avg{\hat{T}^2(z)}_{\textrm{D}} &\equiv 
\int_0^z dz'\frac{\alpha(z')\Delta t^2(z')}{N(z')},
\label{RW}
\end{align}
a less well-known chirp-induced jitter given by the fifth term,
\begin{align}
\Avg{\hat{T}^2(z)}_{\textrm{C}} &\equiv
\int_0^z dz'\beta(z')\int_0^{z'} dz''\frac{\alpha(z'')C(z'')}{N(z'')},
\label{chirp_induced}
\end{align}
and also the Gordon-Haus timing jitter \cite{gordon} given by the
sixth term,
\begin{align}
\Avg{\hat{T}^2(z)}_{\textrm{GH}} &\equiv
2\int_0^z dz'\beta(z')\int_0^{z'} dz''\beta(z'')
\nonumber\\&\quad\times
\int_0^{z''}dz'''\frac{\alpha(z''')\Delta\omega^2(z''')}{N(z''')}.
\label{GH}
\end{align}
In most cases considered here, $N \gg 1$, $ \Avg{\hat{T}^2}\ll \Delta
t^2$, and $\Avg{\hat{\Omega}^2} \ll \Delta\omega^2$, so one can use
the classical nonlinear Schr\"odinger equation, Eq.~(\ref{cnse}), to
predict the evolution of $\Delta t(z)$, $C(z)$, and $\Delta\omega(z)$
accurately.  The evolution of $\Avg{\hat{T}^2(z)}$ can subsequently be
calculated analytically or numerically using Eq.~(\ref{jitter}) and
the classical evolution of $\Delta t(z)$, $C(z)$, and
$\Delta\omega(z)$, analogous to the linearized approach
\cite{haus_josab,haus_josab2}.

It is worth noting that the chirp-induced jitter,
Eq.~(\ref{chirp_induced}), depends on the cross-correlation between
the position and momentum noise in Eq.~(\ref{cross_correlation}),
so it can be positive as well as negative, but the sum of the
three sources of jitter must obviously remain positive.

\section{\label{soliton}Effect of Loss on Adiabatic Soliton Control}
\subsection{Review of the ideal case}
\begin{figure}[htbp]
\includegraphics[width=0.45\textwidth]{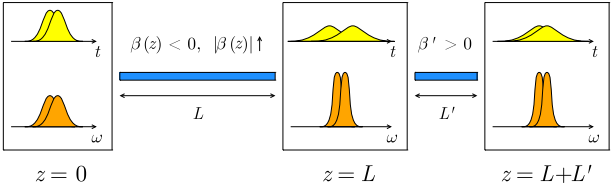}
\caption{(Color online) Schematic (not-to-scale) of the adiabatic
  soliton control scheme. An optical pulse is coupled into a
  dispersion-increasing fiber of length $L$ with a negative
  dispersion coefficient $\beta$, followed by a much shorter
  dispersion-compensating fiber of length $L'$ with a positive
  dispersion coefficient $\beta'$.}
\label{adiabatic_setup}
\end{figure}

Consider the scheme proposed in Ref.~\cite{tsang_prl} and depicted in
Fig.~\ref{adiabatic_setup}. Assume that the dispersion coefficient of
the first fiber $\beta(z)$ is negative and its magnitude increases
along the fiber slowly compared with the soliton period. The classical
soliton solution of Eq.~(\ref{cnse}), assuming adiabatic change in
parameters $\beta(z)$ and $N(z)$, is \cite{kuehl}
\begin{align}
A(z,t) &= A_0(z)\sech\left[ \frac{t}{\tau(z)}\right]
\exp\left[\frac{i\kappa}{2}\int_0^z dz'|A_0(z')|^2\right],
\\
A_0(z) &= \sqrt{\frac{N(z)}{2\tau(z)}},\
\tau(z) =\frac{2|\beta(z)|}{\kappa N(z)}.
\end{align}
The adiabatic approximation is satisfied when
\begin{align}
\left|\frac{\beta(z)}{d\beta(z)/dz}\right| &\ll \Lambda,\
\left|\frac{N(z)}{dN(z)/dz}\right| = \frac{1}{\alpha} \ll \Lambda,
\end{align}
where $\Lambda$ is the soliton period,
\begin{align}
\Lambda(z) &\equiv \frac{\pi}{2}\frac{\tau^2(z)}{|\beta(z)|}.
\end{align}
The root-mean-square pulse width $\Delta t(z)$ and
bandwidth $\Delta\omega(z)$ then become
\begin{align}
\Delta t(z) &= \frac{\pi}{2\sqrt{3}}\tau(z) 
=\frac{\pi}{\sqrt{3}}\frac{|\beta(z)|}{\kappa N(z)},
\label{adiabatic_dt}
\\
\Delta\omega(z) &=\frac{1}{\sqrt{3}\tau(z)} =
\frac{1}{2\sqrt{3}}\frac{\kappa N(z)}{|\beta(z)|}.
\label{adiabatic_domega}
\end{align}
The bandwidth $\Delta\omega(z)$ is thus reduced in the first fiber. If
the second fiber has a positive dispersion coefficient $\beta'$ so
that the net dispersion is zero ($\int_0^L dz \beta(z)+\beta'L' = 0$),
the quantum dispersion effect given by the second and third term on
the right-hand side of Eq.~(\ref{jitter}) can be eliminated.
Furthermore, if $\beta'$ has a very large magnitude compared with
$\beta(z)$ so that the second fiber can be very short compared with
the first fiber, the effective nonlinearity experienced by the pulse
in the second fiber can be neglected, and $\Delta\omega(z)$ remains
essentially constant in the second fiber. In the lossless case, the
final timing jitter $\Avg{\hat{T}^2(L+L')}$ is therefore the same as
the input $\Avg{\hat{T}^2(0)}$, but $\Delta\omega(L+L')$ has been
reduced and the standard quantum limit on $\Avg{\hat{T}^2(L+L')}$,
Eq.~(\ref{position_sql}), is raised.  Provided that the initial timing
jitter of a laser pulse obeys the coherent-field statistics given by
Eq.~(\ref{VarT_CF}), the final timing jitter is
\begin{align}
\Avg{\hat{T}^2(L+L')} =\Avg{\hat{T}^2(0)} = \frac{\Delta t^2(0)}{N}
=\frac{\pi^2}{3}\frac{\beta^2(0)}{\kappa^2 N^3},
\end{align}
while the final standard quantum limit is
\begin{align}
\Avg{\hat{T}^2(L+L')}_{\textrm{SQL}} &= \frac{1}{4N\Delta\omega^2(L+L')}
= \frac{3\beta^2(L)}{\kappa^2 N^3}.
\end{align}
A timing jitter squeezing ratio, analagous to the squeezing ratio
defined by Haus and Lai \cite{haus_josab}, can be defined as
\begin{align}
R &= \frac{\Avg{\hat{T}^2(L+L')}}{\Avg{\hat{T}^2(L+L')}_{\textrm{SQL}}}
=\frac{\pi^2}{9}\frac{\beta^2(0)}{\beta^2(L)}.
\end{align}
The factor of $\pi^2/9$ arises because the initial jitter for a sech
pulse shape is slightly higher than the standard quantum limit given
by Eq.~(\ref{position_sql}) in terms of the bandwidth. As long as
$\beta(L)$ at the end of the first fiber is significantly larger than
the initial value, the timing jitter becomes lower than the raised
standard quantum limit, $R$ becomes smaller than $1$, and quantum
enhancement of position accuracy is accomplished. This semiclassical
analysis is valid in all practical cases, where $N \gg 1$, $R \gg
1/N$, $\Avg{\hat{T}^2}\ll\Delta t^2$, $\Avg{\hat{\Omega}^2} \ll
\Delta\omega^2$, and is consistent with the analysis of exact quantum
soliton theory in Ref.~\cite{tsang_prl}.  $R$ is related to the
quantum enhancement factor $\gamma$ defined in Ref.~\cite{tsang_prl}
by $R = 1/\gamma^2$. The semiclassical analysis is no longer valid
when $R$ is close to the Heisenberg limit $1/N$, but as the next
sections will show, owing to decoherence effects, it is extremely
difficult for the enhancement to get anywhere close to the Heisenberg
limit.

\subsection{\label{numerical}Numerical analysis of a realistic case}
To investigate the impact of noise and the validity of the adiabatic
approximation in practice, a numerical evaluation of $\Delta t(z)$,
$C(z)$, $\Delta\omega(z)$, and $\Avg{\hat{T}^2(z)}$, using
Eqs.~(\ref{cnse}) and (\ref{jitter}) and realistic parameters,
is necessary. $\beta(z)$ is assumed to have the following profile
used in Ref.~\cite{bogatyrev},
\begin{align}
\beta(z) &= \frac{-12.75\ \textrm{ps}^2\textrm{/km}}
{1+(L- z)/L_\beta}.
\end{align}
$L_\beta = 1$ km is used here instead of the $L_\beta = 1/12$ km used
in Ref.~\cite{bogatyrev}, in order to satisfy the adiabatic
approximation for a longer pulse in this example.  Other fiber
parameters are $\alpha = 0.4$ dB/km, $n_2 = 2.6 \times 10^{-16}$
cm$^2$/W, $A_{\textrm{eff}}= 30$ $\mu$m$^2$ \cite{bogatyrev},
$\lambda_0 = 1550$ nm, $\omega_0 = 2\pi c/\lambda_0$, so that $\kappa
= \hbar\omega_0 (\omega_0 n_2/cA_{\textrm{eff}})$. $L$ is assumed to
be $2$ km. A dispersion-compensating fiber with $\beta' = 127.5$
ps$^2$/km, $\alpha = 0.4$ dB/km, $n_2 = 2.7\times 10^{-16}$ cm$^2$/W,
$A_{\textrm{eff}} = 15$ $\mu$m$^2$ \cite{gruner}, and $L'=110$ m is
used in the numerical analysis as the second fiber.  The classical
nonlinear Schr\"odinger equation, Eq.~(\ref{cnse}), is numerically
solved using the Fourier split-step method \cite{agrawal}.  An initial
sech soliton pulse with $\tau(0) = 1$ ps, $N(0) = 1.9\times 10^7$, and
an initial energy of $2.4$ pJ is assumed.

\begin{figure}[htbp]
\includegraphics[width=0.45\textwidth]{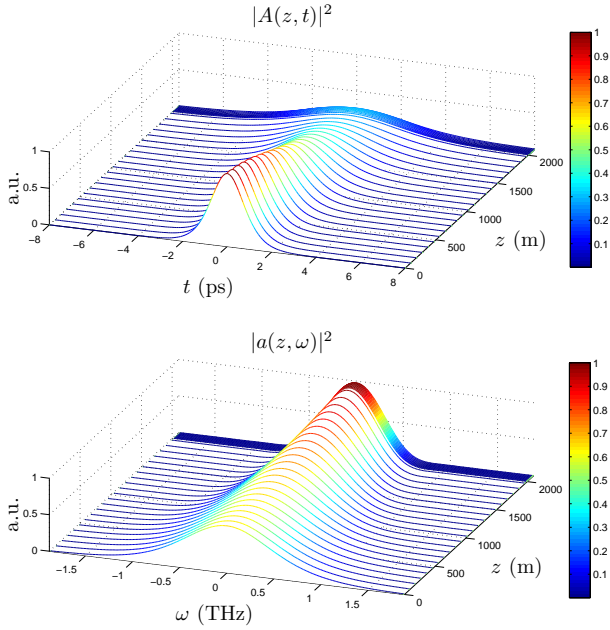}
\caption{(Color online) Numerical evolution of pulse intensity (top)
  and spectrum (bottom).  The denser plots for $z > 2000$ m indicate
  pulse propagation in the second fiber. The color codes are in the same
  arbitrary units as the heights of the plots.}
\label{evolution}
\end{figure}

\begin{figure}[htbp]
\includegraphics[width=0.45\textwidth]{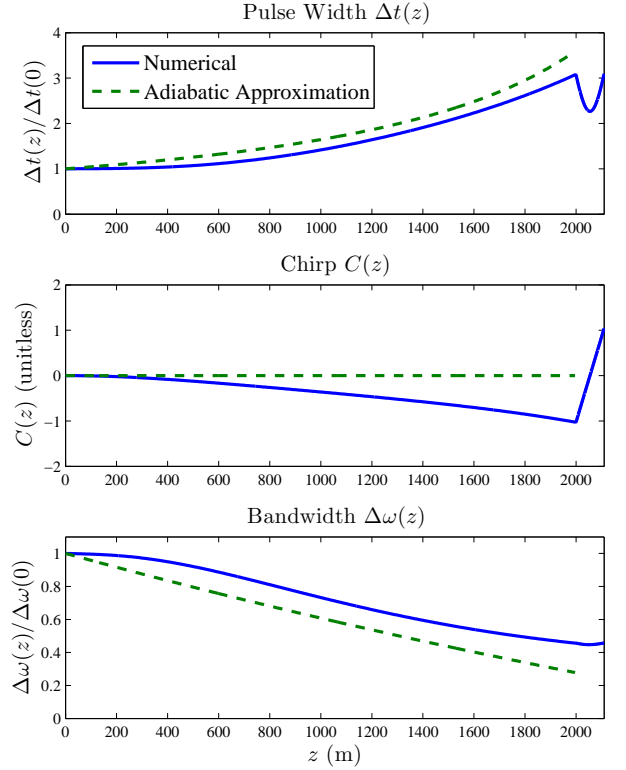}
\caption{(Color online) Evolution of pulse width $\Delta t(z)$ (top),
  chirp $C(z)$ (center), and bandwidth $\Delta\omega$ (bottom),
  compared with the adiabatic approximation (dash).  Plots of $\Delta
  t$ and $\Delta\omega$ are normalized with respect to their initial
  values, respectively.}
\label{pulsewidth_bandwidth}
\end{figure}

Figure \ref{evolution} plots the numerical evolution of pulse
intensity and spectrum in the two fibers. As expected, the bandwidth
is narrowed in the first fiber and remains approximately constant in
the second ($z > 2000$ m), owing to the latter's relative short
length.  Figure \ref{pulsewidth_bandwidth} plots the evolution of
pulse width $\Delta t(z)$, chirp $C(z)$, and bandwidth
$\Delta\omega(z)$, compared with the adiabatic approximation,
Eqs.~(\ref{adiabatic_dt}) and (\ref{adiabatic_domega}). The adiabatic
approximation is evidently not exact, and the pulse acquires a chirp
due to excess dispersion in the first fiber, leading to slight
refocusing in the second fiber. The bandwidth is reduced by a factor
of 2.2, as opposed to the ideal factor of 3.6.

\begin{figure}[htbp]
\includegraphics[width=0.45\textwidth]{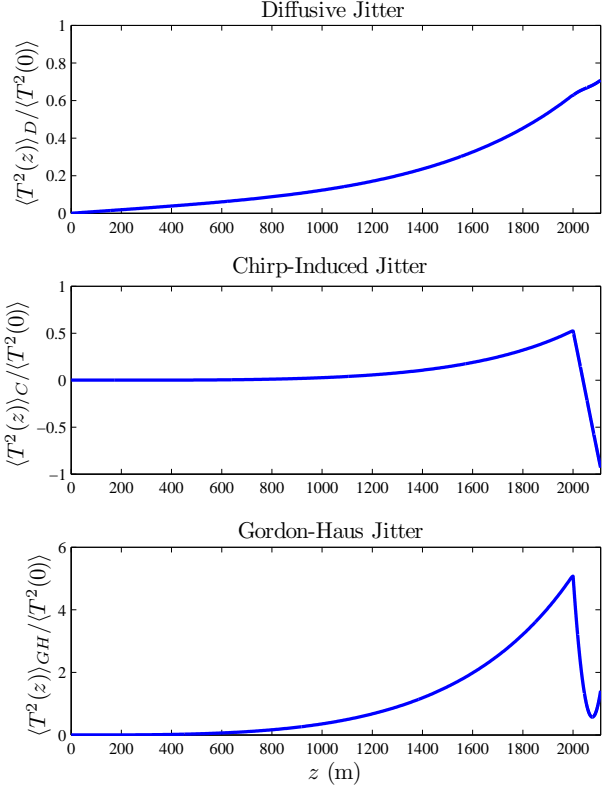}
\caption{(Color online) Evolution of diffusive jitter (top),
  chirp-induced jitter (center), and Gordon-Haus jitter (bottom). All
  plots are normalized with respect to the initial jitter
  $\Avg{\hat{T}^2(0)}$.}
\label{jitters}
\end{figure}

Figure \ref{jitters} plots the evolution of the diffusive jitter
given by Eq.~(\ref{RW}), the chirp-induced jitter given by
Eq.~(\ref{chirp_induced}), and the Gordon-Haus jitter given by
Eq.~(\ref{GH}). It can be seen that although the Gordon-Haus jitter
increases much more quickly than the other jitter components in the
first fiber, the former drops abruptly in the second fiber ($z > 2000$
m) due to the opposite dispersion. This kind of Gordon-Haus jitter
reduction by dispersion management is well known \cite{smith}. The
chirp-induced jitter component drops below zero in the second fiber,
but as noted before, the total noise jitter remains positive.  The
final jitter values are numerically determined to be
$\Avg{\hat{T}^2(L+L')}_{\textrm{D}} = 0.71\Avg{\hat{T}^2(0)}$,
$\Avg{\hat{T}^2(L+L')}_{\textrm{C}} = -0.93\Avg{\hat{T}^2(0)}$, and
$\Avg{\hat{T}^2(L+L')}_{\textrm{GH}} = 1.42\Avg{\hat{T}^2(0)}$, resulting in a
total jitter of
\begin{align}
\Avg{\hat{T}^2(L+L')}&=\Avg{\hat{T}^2(0)}+\Avg{\hat{T}^2(L+L')}_{\textrm{D}}
\nonumber\\&\quad
+\Avg{\hat{T}^2(L+L')}_{C}+\Avg{\hat{T}^2(L+L')}_{\textrm{GH}}
\nonumber\\
&=2.19\Avg{\hat{T}^2(0)}.
\end{align}
The final squeezing ratio is hence
\begin{align}
R &= \frac{\Avg{\hat{T}^2(L+L')}}{\Avg{\hat{T}^2(L+L')}_{\textrm{SQL}}}
\nonumber\\
&=\frac{\pi^2}{9}\frac{\Avg{\hat{T}^2(L+L')}}{\Avg{\hat{T}^2(0)}}
\frac{N(L+L')}{N(0)}
\frac{\Delta\omega^2(L+L')}{\Delta\omega^2(0)}
\nonumber\\
&= 0.42 = -3.8\ \textrm{dB}.
\end{align}
Despite taking into account the increased timing jitter and the
non-ideal bandwidth narrowing, a squeezing ratio of $-3.8$ dB is
predicted by the numerical analysis, suggesting that one should be
able to observe the quantum enhancement experimentally using current
technology.

\subsection{\label{improvements}Potential improvements}
As shown in the previous section, the Gordon-Haus effect contributes
the largest amount of noise in the soliton control scheme, despite its
partial reduction by dispersion management. Its magnitude at the end
of the first fiber can be estimated roughly as
\begin{align}
\frac{\Avg{\hat{T}^2(L)}_{\textrm{GH}}}{\Avg{\hat{T}^2(L)}_{\textrm{SQL}}}
&\sim \left(\frac{L}{\Lambda}\right)^2(\alpha L).
\label{rough_GH}
\end{align}
As the length of the first fiber must be at least a few times longer
than the soliton period $\Lambda$ for the adiabatic approximation to
hold and for the bandwidth to be significantly reduced, $L/\Lambda$ is
approximately fixed, and the Gordon-Haus jitter can be reduced only if
a figure of merit,
\begin{align}
\textrm{FOM} \equiv \frac{1}{\alpha \Lambda} 
= \frac{2}{\pi}\frac{|\beta|}{\alpha\tau^2},
\end{align}
is enhanced. Since this is a rough order-of-magnitude estimate, a
representative value of $\Lambda$, say at $z = L$, can be used. The
figure of merit suggests that the performance of the soliton control
scheme can be improved by reducing the pulse width, increasing the
overall dispersion coefficient, or reducing the loss coefficient.

Reducing the pulse width is the most convenient way of obtaining
better enhancement, as the adiabatic bandwidth reduction can be
achieved over a shorter distance with less loss of photons.  For
example, using $\tau(0) = 500$ fs, $L = 1$ km, $L_\beta = 0.3$ km, $L'
= 44$ m, and otherwise the same parameters as in Sec.~\ref{numerical},
the squeezing ratio becomes $-6.0$ dB, while using $\tau(0) = 200$ fs,
$L = 500$ m, $L_\beta = 1/12$ km, and $L' = 16.2$ m gives a squeezing
ratio of $-7.3$ dB. The shorter pulse width, however, significantly
enhances higher-order dispersive and nonlinear effects.  Raman
scattering, in particular, contributes additional quantum noise
because of coupling to optical phonons \cite{kartner}. It is beyond
the scope of this paper to investigate these higher-order effects, so
a more conservative pulse width of $1$ ps is used in the preceding
section. A larger overall dispersion coefficient, on the other hand,
means that more photons or a higher nonlinearity are required for a
soliton to form, so the Raman effect may also become more significant
with a larger dispersion coefficient. The Raman effect can be reduced
by cooling the fiber and reducing the number of thermal phonons
\cite{kartner}, if it becomes a significant problem.

Further advance in optical fiber technology should be able to increase
the figure of merit by reducing loss, since the specialty fibers
assumed in Sec.~\ref{numerical} have a higher loss than usual
transmission fibers by a factor of two. Using $\alpha = 0.2$ dB/km
instead of $0.4$ dB/km in Sec.~\ref{numerical}, for instance, reduces
the squeezing ratio to $-4.7$ dB. Spectral filtering or
frequency-dependent gain \cite{mecozzi} provides another way of
controlling the Gordon-Haus effect, although it adds another level of
complexity to the experimental setup, and it is beyond the scope of
this paper to investigate how the frequency-dependent dissipation or
amplification might help the quantum enhancement scheme.  Finally, the
design of the setup assumed in Sec.~\ref{numerical} is not fully
optimized, and further optimization of parameters, fiber dispersion
profiles, and bandwidth narrowing strategy should be able to improve
the enhancement.

\section{\label{transmission}Effect of Loss on the Transmission of
  Quantum-Enhanced Timing Information}
Provided that quantum enhancement of pulse position accuracy can be
achieved, the information still needs to be transmitted through
unavoidably lossy channels. It is hence an important question to ask
how loss affects the quantum-enhanced information in optical
information transmission systems. Equation (\ref{jitter})
governs the general evolution of the timing jitter under the effects
of loss, dispersion, and nonlinearity, but in order to estimate the
relative magnitude of the decoherence effects and gain more insight
into the decoherence processes, in this section
Eq.~(\ref{jitter}) is explicitly solved for various systems and
compared with the standard quantum limit.

\subsection{Linear non-dispersive systems}
Without dispersion, the timing jitter increases only due to the
diffusive component $\Avg{\hat{T}^2(z)}_{\textrm{D}}$. An analytic expression for
$\Avg{\hat{T}^2(z)}$ can then be derived from
Eq.~(\ref{jitter}), as $\Delta t(z)$ and $\Delta\omega(z)$
remain constant,
\begin{align}
\Avg{\hat{T}^2(z)} &= \Avg{\hat{T}^2(0)} +\frac{\Delta t^2(0)}{N(z)}
(1-e^{-\alpha z}).
\label{random_walk}
\end{align}
If the initial variance obeys coherent-field statistics, that is,
$\Avg{\hat{T}^2(0)} = \Delta t^2(0)/N(0)$ according to
Eq.~(\ref{VarT_CF}), the subsequent jitter is 
\begin{align}
\Avg{\hat{T}^2(z)}_{\textrm{coh}} = \frac{\Delta t^2(0)}{N(z)},
\end{align}
and obeys the same coherent-field statistics but for the reduced
photon number $N(z)$.  This is consistent with intuition.  On the
other hand, in the high loss limit ($\alpha z \gg 1$), the term
$\Delta t^2(0)/N(z)$ is likely to dominate over the initial jitter
$\Avg{\hat{T}^2(0)}$, so in most cases the position of a significantly
attenuated pulse relaxes to coherent-field statistics independent of
its initial fluctuation.  This justifies the assumption in
Sec.~\ref{soliton} that a laser pulse exiting a laser cavity has such
statistics, regardless of the quantum properties of the pulse inside
the cavity.

Equation (\ref{random_walk}) can be renormalized as
\begin{align}
R(z) &\equiv \frac{\Avg{\hat{T}^2(z)}}{\Avg{\hat{T}^2(z)}_{\textrm{SQL}}}
\nonumber\\
&= R(0)e^{-\alpha z} 
+4\Delta t^2(0)\Delta\omega^2(0)(1-e^{-\alpha z}).
\label{random_walk_norm}
\end{align}
When $\Avg{\hat{T}^2}\ll \Delta t^2$ and $\Avg{\hat{\Omega}^2} \ll
\Delta\omega^2$, classical theory predicts that $4\Delta
t^2\Delta\omega^2 \approx 1$. Equation (\ref{random_walk_norm}) then
suggests that the relative increase in timing jitter is independent of
the initial squeezing ratio $R(0)$. This is nevertheless not true in
general, as $\Delta t$ may depend on both $\Delta\omega$ and $R$ when
the classical theory fails. In Appendix \ref{time_bandwidth}, the
exact dependence of $\Delta t$ on $\Delta\omega$ and $R$ is calculated
for a specific multiphoton state with a Gaussian pulse shape called
the jointly Guassian state. The expression $4\Delta t^2\Delta\omega^2$
is given by
\begin{align}
4\Delta t^2\Delta\omega^2 &= 
\frac{R}{N}+\frac{(1-1/N)^2}{1-1/(NR)},
\label{tbp}
\end{align}
which results in the following exact expression for an initial
jointly Gaussian state,
\begin{align}
R(z) &=R(0)e^{-\alpha z}+
\left[\frac{R}{N}+\frac{(1-1/N)^2}{1-1/(NR)}\right]_{z=0}
(1-e^{-\alpha z}).
\label{random_walk_gauss}
\end{align}
For a large photon number ($N \gg 1$) and moderate enhancement ($1\ge
R \gg 1/N$), $4\Delta t^2\Delta\omega^2\approx1$, as classical theory
would predict for a Gaussian pulse. In this regime, the
quantum-enhanced information is just as sensitive to loss as
standard-quantum-limited information. When $R$ gets close to the
Heisenberg limit $1/N$, however, $\Delta t\Delta\omega$ approaches
infinity.  This is because maximal coincident-frequency correlations
are required to achieve the Heisenberg limit
\cite{giovannetti_nature}, but heuristically speaking, if the photons
have exactly the same momentum, they must have infinite uncertainties
in their relative positions, leading to an infinite pulse width
$\Delta t$.  Owing to the abrupt increase in $4\Delta
t^2\Delta\omega^2$ when $R$ approaches $1/N$, the quantum enhancement
becomes much more sensitive to loss.  In the Heisenberg limit of $R
\to 1/N$, $\Delta t\to \infty$, any loss completely detroys the timing
accuracy and leads to an infinite jitter, according to
Eq.~(\ref{random_walk_gauss}).

\subsection{Linear dispersive systems}
If the system is lossy, dispersive, but linear, it is not difficult
to show that
\begin{align}
\Delta t^2(z) &= \Delta t^2(0) +C(0)\int_0^zdz'\beta(z')
\nonumber\\&\quad
+\Delta\omega^2(0)\left[\int_0^zdz'\beta(z')\right]^2,
\\
C(z) &= C(0) + 2\Delta\omega^2(0)\int_0^zdz'\beta(z'),
\\
\Delta\omega^2(z) &= \Delta\omega^2(0).
\end{align}
The following result can then be obtained from
Eq.~(\ref{jitter}) after some algebra,
\begin{align}
\Avg{\hat{T}^2(z)} &=\Avg{\hat{T}^2(0)}+
\Avg{\hat{T}(0)\hat{\Omega}(0)+\hat{\Omega}(0)\hat{T}(0)}
\int_0^zdz'\beta(z')
\nonumber\\&\quad
+\Avg{\hat{\Omega}^2(0)}\left[\int_0^z dz'\beta(z')\right]^2
+\frac{\Delta t^2(z)}{N(z)}(1-e^{-\alpha z}).
\label{dispersive_jitter}
\end{align}
This result is similar to that in the previous section, except for the
presence of quantum dispersion and the dispersive spread of the pulse
width $\Delta t(z)$ that leads to increased jitter.  With initially
coherent-field statistics, $\Avg{\hat{T}^2(0)}$ and
$\Avg{\hat{\Omega}^2(0)}$ are given by Eqs.~(\ref{VarT_CF}) and
(\ref{VarOmega_CF}), respectively, while by similar arguments, the
coherent-field statistics for
$\Avg{\hat{T}\hat{\Omega}+\hat{\Omega}\hat{T}}$ is
\begin{align}
\Avg{\hat{T}\hat{\Omega}+\hat{\Omega}\hat{T}}_{\textrm{coh}}
&=\frac{C}{N}.
\end{align}
This leads to the following position variance for a pulse with
initially coherent-field statistics,
\begin{align}
\Avg{\hat{T}^2(z)}_{\textrm{coh}} &= \frac{\Delta t^2(z)}{N(z)},
\end{align}
which still maintains the coherent-field statistics for the dispersed
pulse width and the reduced photon number. In the high loss limit
($\alpha z \gg 1$), the coherent-field statistics is again approached
regardless of the initial conditions.

For an initial jointly Gaussian quantum state, on the other hand,
the normalized version of Eq.~(\ref{dispersive_jitter}) is
\begin{align}
R(z) &=\left[R(0)+\frac{4}{R(0)}\zeta^2\right]e^{-\alpha z}
\nonumber\\&\quad
+\left[4\Delta t^2(0)\Delta\omega^2(0)+4\zeta^2\right](1-e^{-\alpha z}),
\end{align}
where $\zeta$ is the normalized effective propagation distance,
\begin{align}
\zeta &\equiv \Delta\omega^2(0)\int_0^z dz'\beta(z'),
\end{align}
and $4\Delta t^2(0)\Delta\omega^2(0)$ is given by Eq.~(\ref{tbp})
evaluated at $z = 0$. As long as the loss is moderate so that
$e^{-\alpha z} \gg 1-e^{-\alpha z}$, quantum dispersion, given by the
term proportional $\zeta^2/R(0)$, becomes the dominant effect and
overwhelms the initial enhancement when $\zeta$ exceeds $R(0)/2$.

If the net dispersion $\int_0^z dz'\beta(z')$ is zero, both quantum
and classical dispersion are eliminated, and the jitter growth becomes
identical to that in a non-dispersive and linear system given by
Eq.~(\ref{random_walk}).

\subsection{Soliton-like systems}
The previous sections show that coherent-field statistics is maintained in
a linear system, but as Sec.~\ref{soliton} clearly shows,
non-trivial statistics can arise from the quantum dynamics of a
nonlinear system. The complex evolution of $\Delta t(z)$, $C(z)$, and
$\Delta\omega(z)$ in general prevents one from solving
Eq.~(\ref{jitter}) explicitly, except for special cases such as
solitons.

If the dispersion is constant and the pulse propagates in the fiber as
a soliton, $C(z)$ is zero, while $\Delta t(z)$ and $\Delta\omega(z)$
can be regarded as constant if $\Avg{\hat{T}^2}\ll\Delta t^2$ and
$\Avg{\hat\Omega^2}\ll \Delta\omega^2$ throughout propagation.
Equation (\ref{jitter}) can then be solved explicitly,
\begin{align}
\Avg{\hat{T}^2(z)} &= \Avg{\hat{T}^2(0)}+
\Avg{\hat{\Omega}^2(0)}\beta^2z^2
+\frac{\Delta t^2(0)}{N(0)}(e^{\alpha z}-1)
\nonumber\\&\quad
+\frac{2\beta^2\Delta\omega^2(0)}{N(0)}
\left(\frac{e^{\alpha z}-1}{\alpha^2}-\frac{z}{\alpha}-\frac{z^2}{2}\right),
\label{constant_disp}
\end{align}
where $\Avg{\hat{T}(0)\hat{\Omega}(0)+\hat{\Omega}(0)\hat{T}(0)}$ is
asssumed to be zero for simplicity. If
$4\Avg{\hat{T}^2(0)}\Avg{\hat{\Omega}^2(0)} =(\pi^2/9)[1/N^2(0)]$ is
also assumed for a soliton pulse for simplicity,
Eq.~(\ref{constant_disp}) can be normalized to give
\begin{align}
R(z) &= R(0)e^{-\alpha z}+
\frac{\pi^4}{81}\frac{1}{R(0)}\left(\frac{z}{\Lambda}\right)^2e^{-\alpha z}
+\frac{\pi^2}{12}\frac{\tau^2}{N(0)}(e^{\alpha z}-1)
\nonumber\\&\quad
+\frac{2\pi^2}{9}\frac{e^{-\alpha z}}{\Lambda^2}
\left(\frac{e^{\alpha z}-1}{\alpha^2}-\frac{z}{\alpha}-\frac{z^2}{2}\right).
\label{normalized_R}
\end{align}
In the low loss regime with $\alpha\Lambda \ll 1$ and $\alpha z \ll
1$, Eq.~(\ref{normalized_R}) can be further simplified,
\begin{align}
R(z) &\approx R(0) + 
\frac{\pi^4}{81}\frac{1}{R(0)}\left(\frac{z}{\Lambda}\right)^2
+\frac{\pi^2}{12}\frac{\tau^2}{N(0)}(\alpha z)
\nonumber\\&\quad
+\frac{\pi^2}{27}\left(\frac{z}{\Lambda}\right)^2(\alpha z).
\label{simp_norm_R}
\end{align}
Quantum dispersion is again the dominant effect in this regime, while
decoherence effects are much smaller, by a factor of $\alpha z$
approximately.

Even if the net dispersion is zero and quantum dispersion is
compensated, the Gordon-Haus effect cannot be completely eliminated
by dispersion management in the presence of nonlinearity and may
become significant, as the numerical analysis in Sec.~\ref{numerical}
shows.  An order-of-magnitude estimate of Gordon-Haus jitter can be
performed by considering soliton propagation in a constant negative
dispersion fiber, just as in the previous case, followed by a
dispersion-compensating fiber of length $L'$ with positive dispersion
coefficient $\beta'$. If $L'$ is short, the effective nonlinearity
experienced by the pulse in the second fiber can be neglected, and
$\Delta\omega(z)$ can be regarded as constant. Assuming that $\beta L
+ \beta'L' = 0$, the integral in Eq.~(\ref{GH}) can be solved
to give the Gordon-Haus jitter,
\begin{align}
\Avg{\hat{T}^2(L+L')}_{\textrm{GH}} &\approx
\frac{\alpha\Delta\omega^2(0)}{6N(0)}\beta^2L^2(L+L')
\nonumber\\
&\approx \frac{\alpha\Delta\omega^2(0)}{6N(0)}\beta^2L^3.
\end{align}
The normalized contribution to the squeezing ratio is therefore
\begin{align}
\frac{\Avg{\hat{T}^2(L+L')}_{\textrm{GH}}}{\Avg{\hat{T}^2(L+L')}_{\textrm{SQL}}}
&\approx \frac{\pi^2}{54}\left(\frac{L}{\Lambda}\right)^2 (\alpha L).
\end{align}
Compared with the Gordon-Haus jitter at the end of the first fiber
given by the last term of Eq.~(\ref{simp_norm_R}), dispersion
management cuts the jitter by half, but the expression maintains its
functional dependence on the parameters of the first fiber.  This
estimate also justifies the use of Eq.~(\ref{rough_GH}) to estimate
the Gordon-Haus jitter at the end of the two fibers in
Sec.~\ref{improvements}.  To minimize the impact of Gordon-Haus jitter
on the quantum-enhanced timing accuracy in a dispersion-managed
soliton system, the condition
\begin{align}
L^3 \ll \frac{54}{\pi^2}\left(\frac{\Lambda^2}{\alpha}\right)R
\end{align}
is required.

\section{Conclusion}
In conclusion, the decoherence effect by optical loss on adiabatic
soliton control and on the transmission of quantum-enhanced timing
information has been extensively studied. It is found that an
appreciable enhancement can still be achieved by the soliton scheme
using current technology, despite an increase of timing jitter due to
the presence of loss. It is also found that the quantum-enhanced
timing accuracy should be much lower than the Heisenberg-limited
accuracy to avoid increased sensitivity to photon loss during
transmission, and the net dispersion in the transmission system should
be minimized in order to reduce quantum dispersion and the Gordon-Haus
effect.

Although the most important pulse propagation effects have been
considered in this analysis, higher-order effects, such as third-order
dispersion, self-steepening, and Raman scattering \cite{agrawal} might
provide further adverse impact on the quantum enhancement if the
optical pulse is ultrashort. In particular, the inelastic Raman
scattering process is expected to be a significant source of
decoherence for ultrashort pulses \cite{kartner}.  It is beyond the
scope of this paper to investigate these higher-order effects, but
they should be of minor importance for picosecond pulses and the
propagation distances considered in this paper.

Finally, it is worth noting that while this paper focuses on optical
pulses, the developed formalism is equally valid for describing the
transverse position and momentum of optical beams \cite{beam} and the
center-of-mass variables of Bose-Einstein condensates \cite{vaughan}.
Decoherence by loss of particles in those systems can be studied using
the formalism developed in this paper and parameters specific to those
systems.

\section{Acknowledgments}
This work is financially supported by the DARPA Center for Optofluidic
Integration and the National Science Foundation through the Center for
the Science and Engineering of Materials (DMR-0520565).
\appendix

\section{\label{exact_heisenberg}Derivation of Exact Heisenberg Limits}
An exact Heisenberg limit can be derived if the inverse
photon-number operator $\hat{N}^{-1}$ is used instead of $1/N$ in the
definitions of $\hat{T}$, $\hat{\Omega}$, $\Delta t$, and
$\Delta\omega$ in Eqs.~(\ref{position}), (\ref{momentum}),
(\ref{pulsewidth}), and (\ref{bandwidth}), just as
in Refs.~ \cite{vaughan} and \cite{lai1},
\begin{align}
\hat{T}' &\equiv \hat{N}^{-1}\int dt\ t\hat{A}^\dagger\hat{A},
\\
\hat{\Omega}' &\equiv
\hat{N}^{-1}\int d\omega\ \omega\hat{a}^\dagger\hat{a},
\\
\Delta t' &\equiv
\Avg{\hat{N}^{-1}\int dt\ t^2\hat{A}^\dagger\hat{A}}^{1/2},
\\
\Delta\omega' &\equiv
\Avg{\hat{N}^{-1}\int d\omega\ \omega^2\hat{a}^\dagger\hat{a}}^{1/2}.
\end{align}
These operators are well defined as long as the quantum state has zero
vacuum-state component ($\langle 0|\hat{\rho}\ket{0} = 0$).
Starting from the Heisenberg uncertainty principle for $\hat{T}'$ and
$\hat{\Omega}'$,
\begin{align}
\Avg{\hat{T}'^2}\Avg{\hat{\Omega}'^2} \ge \frac{\Avg{\hat{N}^{-2}}}{4},
\end{align}
and the inequality
\begin{align}
\Avg{\hat{N}^{-2}\int d\omega \int d\omega'\
(\omega-\omega')^2\hat{a}^\dagger\hat{a}'^\dagger\hat{a}\hat{a}'}\ge 0,
\end{align}
one can obtain the exact inequality
\begin{align}
\Avg{\hat{\Omega}'^2} \le  \Delta\omega'^2,
\end{align}
and the exact Heisenberg limit for the new position operator,
\begin{align}
\Avg{\hat{T}'^2}_{\textrm{H}} &= \frac{\Avg{\hat{N}^{-2}}}{4\Delta\omega'^2}.
\label{exact_h}
\end{align}
The difference between Eqs.~(\ref{h_limit}) and
(\ref{exact_h}) is negligible for small photon-number
fluctuations.  The exact Heisenberg limit on $\Avg{\hat{\Omega}'^2}$
is similar.

\section{\label{langevin}Noise Statistics}
In this section the expression $\Avg{\hat{S}_T(z)\hat{S}_T(z')}$ in
Eq.~(\ref{position_noise}) is calculated. The derivations of
$\Avg{\hat{S}_\Omega(z)\hat{S}_\Omega(z')}$ in
Eq.~(\ref{momentum_noise}) and $\Avg{\hat{S}_T(z)\hat{S}_\Omega(z')
  +\hat{S}_\Omega(z)\hat{S}_T(z')}$ in Eq.~(\ref{cross_correlation})
are similar. Substituting Eq.~(\ref{S_T}) into
Eq.~(\ref{position_noise}) gives
\begin{align}
\Avg{\hat{S}_T(z)\hat{S}_T(z')}
&=\frac{1}{NN'}\int dt \int dt'\
tt'\Big[\Avg{\hat{s}^\dagger \hat{A}\hat{s}'^\dagger\hat{A}'}
+\Avg{\hat{s}^\dagger\hat{A}\hat{A}'^\dagger\hat{s}'}
\nonumber\\&\quad 
+\Avg{\hat{A}^\dagger\hat{s}\hat{A}'^\dagger\hat{s}'}
+\Avg{\hat{A}^\dagger\hat{s}\hat{s}'^\dagger\hat{A}'}
\Big],
\label{S_T_explicit}
\end{align}
where $N = N(z)$, $N' = N(z')$, $\hat{s} = \hat{s}(z,t)$, $\hat{A} =
\hat{A}(z,t)$, $\hat{s}' = \hat{s}(z',t')$, and $\hat{A}' =
\hat{A}(z',t')$. If the noise reservoir is in the vacuum state,
$\hat{s}\ket{0_{\textrm{reservoir}}} =
\bra{0_{\textrm{reservoir}}}\hat{s}^\dagger = 0$, so only the last
term in Eq.~(\ref{S_T_explicit}) is non-zero,
\begin{align}
\Avg{\hat{S}_T(z)\hat{S}_T(z')} &=
\frac{1}{NN'}\int dt \int dt'\ tt'
\Avg{\hat{A}^\dagger\hat{s}\hat{s}'^\dagger\hat{A}'}
\nonumber\\
&=\frac{1}{NN'}\int dt \int dt'\ tt'
\nonumber\\&\quad\times\bigg[\Avg{\hat{A}^\dagger\hat{A}'}
\alpha\delta(z-z')\delta(t-t')
+\Avg{\hat{A}^\dagger\hat{s}'^\dagger\hat{s}\hat{A}'}\bigg]
\nonumber\\
&=\frac{\alpha \Delta t^2}{N}\delta(z-z')
\nonumber\\&\quad
+\frac{1}{NN'}\int dt \int dt'\ tt'
\Avg{\hat{A}^\dagger\hat{s}'^\dagger\hat{s}\hat{A}'}.
\label{S_T_explicit2}
\end{align}
The first term on the right-hand side of Eq.~(\ref{S_T_explicit2}) is
the desired result, while the second term can be rewritten as
\begin{align}
&\quad \frac{1}{NN'}\int dt \int dt'\ tt'
\Avg{\hat{A}^\dagger\hat{s}'^\dagger\hat{s}\hat{A}'}
\nonumber\\
&=\frac{1}{NN'}\int dt \int dt'\ tt'
\Avg{\left[\hat{A}^\dagger,\hat{s}'^\dagger\right]
\left[\hat{s},\hat{A}'\right]}.
\label{S_T_explicit3}
\end{align}
If the system is linear, the commutator between $\hat{s}$ and
$\hat{A}$ is always zero \cite{mandel}, but because $\hat{s}$ does not
commute with $\hat{A}^\dagger$ and $\hat{A}$ is coupled to
$\hat{A}^\dagger$ by the nonlinear term in Eq.~(\ref{qnse}), $\hat{s}$
may fail to commute with $\hat{A}$. That said, it can be argued that
the optical field operator must always commute with future noise
operators due to causality and the infinitesimally short memory of
$\hat{s}$,
\begin{align}
\left[\hat{A}^\dagger,\hat{s}'^\dagger\right]&= 0\ \textrm{if}\ z<z',
\\\left[\hat{s},\hat{A}'\right]&= 0\ \textrm{if}\ z>z',
\end{align}
so Eq.~(\ref{S_T_explicit3}) can be non-zero only at $z =
z'$.  The commutator between $\hat{s}$ and $\hat{A}$ at $z = z'$ due
to the parametric coupling of $\hat{A}$ and $\hat{A}^\dagger$ can be
estimated by a perturbative technique.  Consider an integral form of
Eq.~(\ref{qnse}) with the nonlinear term and the Langevin noise term
only,
\begin{align}
\hat{A}(z+\Delta z) &= \hat{A}(z)+
\int_{z}^{z+\Delta z} dz'
\left[i\kappa \hat{A}^\dagger(z')\hat{A}(z')\hat{A}(z')
+ \hat{s}(z')\right],
\label{integral_nls}
\end{align}
and $\hat{A}^\dagger(z')$ given by the Hermitian conjugate of
Eq.~(\ref{integral_nls}),
\begin{align}
\hat{A}^\dagger(z') &= \hat{A}^\dagger(z)+
\int_z^{z'} dz''\left[-i\kappa \hat{A}^\dagger(z'')\hat{A}^\dagger(z'')
\hat{A}(z'')+ \hat{s}^\dagger(z'')\right].
\label{integral_nls_conj}
\end{align}
The commutator between $\hat{s}$ and $\hat{A}$ at $z +\Delta z$ becomes
\begin{align}
&\quad
[\hat{s}(z+\Delta z), \hat{A}(z+\Delta z)]
\nonumber\\&
=i\kappa\int_{z}^{z+\Delta z} dz'
\left[\hat{s}(z+\Delta z),\hat{A}^\dagger(z')\right]\hat{A}(z')\hat{A}(z').
\end{align}
$\hat{s}(z+\Delta z)$ commutes with $\hat{A}(z')$ because $z+\Delta z
> z'$, while it fails to commute with $\hat{A}^\dagger(z')$ because
$\hat{A}^\dagger(z')$ given by Eq.~(\ref{integral_nls_conj}) depends
explicitly on $\hat{s}^\dagger$. Thus, in the leading order of $\Delta
z$,
\begin{align}
&\quad
[\hat{s}(z+\Delta z),\hat{A}(z+\Delta z)]
\nonumber\\&
\approx i\kappa \int_{z}^{z+\Delta z}dz'\int_z^{z'}dz''
\left[\hat{s}(z+\Delta z),\hat{s}^\dagger(z'')\right]\hat{A}(z')\hat{A}(z'),
\end{align}
which approaches $0$ in the limit of $\Delta z \to 0$. Hence $\hat{s}$
commutes with $\hat{A}$ at $z = z'$, and the position noise is given
only by the first term on the right-hand side of
Eq.~(\ref{S_T_explicit2}), resulting in Eq.~(\ref{position_noise}).

\section{\label{time_bandwidth}The Jointly Gaussian State}
A Fock state can be expressed as \cite{mandel,tsang_pra}
\begin{align}
\ket{N} &= \int d\omega_1\ldots\int d\omega_N\ \phi(\omega_1,\ldots,\omega_N)
\ket{\omega_1,\ldots,\omega_N},
\nonumber\\
&= \int dt_1\ldots\int dt_N\ \psi(t_1,\ldots,t_N)\ket{t_1,\ldots,t_N},
\end{align}
where the spectral and temporal eigenstates are given by
\begin{align}
\ket{\omega_1,\ldots,\omega_N} &\equiv
\frac{1}{\sqrt{N!}}\hat{a}^\dagger(\omega_1)\ldots\hat{a}^\dagger(\omega_N)\ket{0},
\\
\ket{t_1,\ldots,t_N} &\equiv
\frac{1}{\sqrt{N!}}\hat{A}^\dagger(t_1)\ldots\hat{A}^\dagger(t_N)\ket{0}.
\end{align}
Theses states are eigenstates of the following operators relevant
to our purpose,
\begin{align}
\hat{\Omega}\ket{\omega_1,\ldots,\omega_N} &=
\bigg(\frac{1}{N}\sum_{n=1}^N\omega_n\bigg)\ket{\omega_1,\ldots,\omega_N},
\label{config_omega}\\
\hat{T}\ket{t_1,\ldots,t_N} &=
\bigg(\frac{1}{N}\sum_{n=1}^N t_n\bigg)\ket{t_1,\ldots,t_N},
\label{config_t}\\
\frac{1}{N}\int d\omega\ \omega^2\hat{a}^\dagger\hat{a}
\ket{\omega_1,\ldots,\omega_N} &=
\bigg(\frac{1}{N}\sum_{n=1}^N\omega_n^2\bigg)
\ket{\omega_1,\ldots,\omega_N},
\label{config_domega2}\\
\frac{1}{N}\int dt\ t^2\hat{A}^\dagger\hat{A}\ket{t_1,\ldots,t_N} &=
\bigg(\frac{1}{N}\sum_{n=1}^N t_n^2\bigg)\ket{t_1,\ldots,t_N}.
\label{config_dt2}
\end{align}
$\phi(\omega_1,\ldots,\omega_N)$ is the spectral multiphoton probability
amplitude, and it is related to the temporal probability amplitude
$\psi(t_1,\ldots,t_N)$ by the $N$-dimensional Fourier transform in the
slowly-varying envelope regime. Both amplitudes should also satisfy
normalization and boson symmetry.  To study temporal quantum
enhancement, it is convenient to define the probability amplitude as a
jointly Gaussian function \cite{tsang_pra},
\begin{align}
\phi(\omega_1,\ldots,\omega_N) &= C
\exp\bigg[-\frac{1}{4B^2}\bigg(\frac{1}{N}\sum_{n=1}^N\omega_n\bigg)^2
\nonumber\\&\quad
-\frac{1}{4b^2}\sum_{n=1}^N
\bigg(\omega_n-\frac{1}{N}\sum_{m=1}^N\omega_m\bigg)^2\bigg],
\\
\psi(t_1,\ldots,t_N) &= C'
\exp\bigg[-N^2B^2\bigg(\frac{1}{N}\sum_{n=1}^N t_n\bigg)^2
\nonumber\\&\quad
-b^2\sum_{n=1}^N
\bigg(t_n-\frac{1}{N}\sum_{m=1}^Nt_m\bigg)^2\bigg],
\end{align}
where $B$ and $b$ are arbitrary and real constants, and $C$ and $C'$
are normalization constants. Explicit expressions for
$\Avg{\hat{\Omega}^2}$, $\Avg{\hat{T}^2}$, $\Delta\omega^2$, and
$\Delta t^2$ can be obtained using
Eqs.~(\ref{config_omega})-(\ref{config_dt2}) and Appendix B of
Ref.~\cite{tsang_pra},
\begin{align}
\Avg{\hat{\Omega}^2} &= B^2,
\\
\Avg{\hat{T}^2} &=\frac{1}{4N^2B^2},
\\
\Delta\omega^2 &= B^2 + \left(1-\frac{1}{N}\right)b^2,
\label{dw}
\\
\Delta t^2 &= \frac{1}{4N^2B^2} + \left(1-\frac{1}{N}\right)\frac{1}{4b^2}.
\label{dt}
\end{align}
In the limit of $b \to 0$, $\Avg{\hat{T}^2}$ reaches the Heisenberg
limit,
\begin{align}
\Avg{\hat{T}^2} &= \frac{1}{4N^2\Delta\omega^2},
\end{align}
and the quantum state can be written as a state of 
photons with maximal coincident-frequency correlations,
\begin{align}
\ket{N} &\propto \int d\omega\ \exp\left(-\frac{\omega^2}{4B^2} \right)
\ket{\omega,\ldots,\omega}.
\end{align}
On the other hand, when $B^2 = b^2/N$, $\Avg{\hat{T}^2}$ is at the
standard quantum limit,
\begin{align}
\Avg{\hat{T}^2} &= \frac{1}{4N\Delta\omega^2},
\end{align}
the quantum state has only one excited Gaussian mode \cite{tsang_pra},
\begin{align}
\ket{N} &\propto \int d\omega_1 \ldots\int d\omega_N\ 
\prod_{n=1}^N\exp\left(-\frac{\omega_n}{4b^2}\right)\ket{\omega_1,\ldots,\omega_N}
\nonumber\\
&\propto\left[\int d\omega\
\exp\left(-\frac{\omega}{4b^2}\right)\hat{a}^\dagger(\omega)\right]^N\ket{0},
\end{align}
and therefore also satisfies the coherent-field statistics
\cite{titulaer,mandel}.  These limits and the corresponding quantum
states are consistent with those suggested in
Ref.~\cite{giovannetti_nature}.  With Eqs.~(\ref{dw}) and (\ref{dt}),
the pulse width $\Delta t$ can be determined explicitly in terms of
$\Delta\omega$ and the squeezing ratio $R=\Delta\omega^2/(NB^2)$,
\begin{align}
\Delta t^2 &= \frac{1}{4\Delta\omega^2}
\left[\frac{R}{N} + \frac{(1-1/N)^2}{1-1/(NR)}\right].
\end{align}

\end{document}